\documentclass[showpacs,prl,twocolumn,aps,superscriptaddress,preprintnumbers,letterpaper]{revtex4}
\usepackage{amsmath,amssymb}
\usepackage{epsfig}
\usepackage{graphicx}
\usepackage{amsmath}
\usepackage{amsfonts}
\usepackage{epstopdf}
\def\slashchar#1{\setbox0=\hbox{$#1$}     		
   \dimen0=\wd0                                 	
   \setbox1=\hbox{/} \dimen1=\wd1               	
   \ifdim\dimen0>\dimen1                        	
      \rlap{\hbox to \dimen0{\hfil/\hfil}}      	
      #1                                        	
   \else                                        	
      \rlap{\hbox to \dimen1{\hfil$#1$\hfil}}   	
      /                                         	
   \fi}

\newcommand{\be}{\begin{equation}}
\newcommand{\ee}{\end{equation}}
\newcommand{\bear}{\begin{eqnarray}}
\newcommand{\eear}{\end{eqnarray}}
\newcommand{\ba}{\begin{array}}
\newcommand{\ea}{\end{array}}

\begin{document}

\title{LPM effect as the origin of the jet fragmentation scaling in heavy ion collisions}

\author{Frash\"er Loshaj}
\affiliation{Department of Physics and Astronomy, Stony Brook University, Stony Brook, New York 11794-3800, USA}

\author{Dmitri E. Kharzeev}
\affiliation{Department of Physics and Astronomy, Stony Brook University, Stony Brook, New York 11794-3800, USA}
\affiliation{Department of Physics,
Brookhaven National Laboratory, Upton, New York 11973-5000, USA}

\date{\today}

\begin{abstract}
We address a recent puzzling result from the LHC: the jet fragmentation functions measured  in $PbPb$ and $pp$ collisions appear very similar in spite of a large medium-induced energy loss (we will call this ``jet fragmentation scaling", JFS).  To model the real-time non-perturbative effects in the propagation of a high energy jet through the strongly coupled QCD matter, we adopt an effective dimensionally reduced description in terms of the  $(1+1)$ quasi-Abelian Schwinger theory. This theory is exactly soluble at any value of the coupling and shares with QCD the properties of dynamical generation of ``mesons" with a finite mass and the screening of ``quark" charge  that are crucial for describing the transition of the jet into hadrons. We find that this approach describes quite well the vacuum jet fragmentation in $e^+e^-$ annihilation 
at $z\geq0.2$ at jet energies in the range of the LHC heavy ion measurements ($z$ is the ratio of hadron and jet momenta). In QCD medium, we find that the JFS is reproduced if the mean free path $\lambda$ of the jet is short, $\lambda \leq 0.3$ fm, which is in accord with the small shear viscosity inferred from the measurements of the collective flow. The JFS holds since at short mean free path the quantum interference (analogous to the Landau-Pomeranchuk-Migdal effect in QED) causes the produced mesons to have low momenta $p \sim m$, where $m \simeq 0.6$ GeV is the typical meson mass.  Meanwhile the induced jet energy loss at short mean free path is much larger than naively expected in string models.
\end{abstract}

\maketitle

Recently, the CMS Collaboration at the LHC reported the measurement of the fragmentation function of jets produced in Pb-Pb collisions at $\sqrt{s} = 2.76$ TeV per nucleon pair \cite{Collaboration:2011cs}. Surprisingly, in spite of the large energy loss signaled by the striking imbalance of jet transverse energies in the di-jet events \cite{Aad:2010bu,Chatrchyan:2011sx} and by the suppression of the large transverse momentum hadrons \cite{Aamodt:2010jd}, it has been found that the jet fragmentation functions in Pb-Pb collisions are very similar to the ones measured in pp collisions. This is true even in the most central Pb-Pb collisions, and for the event classes where the imbalance of the jet energies is the largest \cite{Collaboration:2011cs}. It appears that the jet shape modification takes place only at small transverse momenta of the produced hadrons $p_T \leq 2$ GeV, and at large angles outside of the jet cone  \cite{Collaboration:2011cs}. This result is surprising because the multiple scattering of the jet in the medium does not seem to produce extra high momentum hadrons. 

Let us briefly recall the space-time picture of jet evolution in QCD, see e.g. \cite{Dokshitzer:1991wu}. 
The produced high transverse momentum partons are in general 
far off mass shell and evolve emitting gluons and quark-antiquark pairs; this evolution towards smaller parton virtualities is governed by the QCD renormalization group and described by DGLAP equations. At some scale $Q_0^2 \sim 1 - 3\ {\rm GeV}^2$ the non-perturbative effects of confinement set in and transform the radiated partons into the observed hadrons. The uncertainty principle tells us that this ``hadronization" occurs at longitudinal distance $L_h \simeq z P_{jet}/Q_0^2$, where $z$ is the fraction of the jet transverse momentum  $P_{jet}$ carried by the parton with virtuality $Q_0$. 
Using for the sake of an estimate the values $z=0.2$ 
, $P_{jet} = 100$ GeV and $Q_0 = 2$ GeV$^2$, we get $L_h \simeq 2$ fm. In Pb-Pb collisions, this estimate suggests that the jet evolves down to the scales at which the dynamics becomes non-perturbative well within the produced medium.  

It is widely believed that the quark-gluon plasma at temperatures $T\leq (2 - 3) T_c$ produced at RHIC and LHC is non-perturbative and strongly coupled at scales of the order of 1 GeV.
Combined with our estimate of the jet formation time, this suggests that to understand the CMS result one has to develop an approach to the jet interactions in the medium and its subsequent hadronization that is i) valid at strong coupling; ii) describes properly the transformation of partons into the measured hadrons. 
Within the perturbation theory, the approach to the propagation of the jet in the medium has been developed in \cite{Baier:1996kr,Gyulassy:2000er}; see \cite{Baier:2000mf,Wang:2004dn} for reviews.  It has been established that the Landau-Pomeranchuk-Migdal (LPM) effect \cite{Landau:1953um,Migdal:1956tc,Ter-M} -- the quantum interference of the radiation processes in the interactions with multiple scattering centers in the medium --  is as important in QCD as it was originally found in QED, although the non-Abelian effects modify  the radiation pattern. While the LPM effect has been traditionally treated within the perturbation theory, it can be expected to affect the radiation amplitudes  also at strong coupling. Indeed, the LPM effect may be viewed as a consequence of quantum mechanics and the low-energy Low theorem that is based only on the symmetries of the theory (gauge invariance and the conservation of vector current) and is valid even when the perturbation theory does not apply.

To proceed further we need an effective dynamical theory that can be treated at strong coupling and that can describe the transformation of partons into the color-neutral hadrons.
In spite of the lack of a complete theory of confinement much progress has been made in a qualitative understanding of its possible key ingredients. In particular, it is likely that 
the ``dual Meissner effect" proposed by 't Hooft \cite{'tHooft:1977hy} and Mandelstam \cite{Mandelstam:1974pi} describes correctly the qualitative picture of the confinement of color electric flux. In this picture,  the QCD vacuum contains a 
condensate of magnetic monopoles, and a quasi-Abelian Higgs phenomenon takes place. 

It has been suggested long time ago to use the $(1+1)$ QED (the Schwinger model) \cite{Schwinger:1962tp,Lowenstein:1971fc,Coleman:1975pw} as an effective theory capable to model the transformation of partons into hadrons at high energies \cite{:1974cks}, see also \cite{Fujita:1989vv,Wong:1991ub}. Indeed, the high energy justifies the dimensional reduction to $(1+1)$ dimensions, and the Schwinger model with massless fermions shares many key properties with QCD, including: i) the Higgs phenomenon (local electric charge conservation is spontaneously broken); ii) the spontaneous breaking of global chiral symmetry; iii) the screening of color charge 
(similar to the scenario of confinement for QCD with light quarks proposed by Gribov \cite{Gribov:1999ui}, see \cite{Dokshitzer:2004ie} for a review); iv) axial anomaly and  the $\theta$-vacuum. While the Schwinger model is Abelian, it may model the quasi-Abelian dynamics of QCD emerging due to the condensation of magnetic monopoles in the vacuum that may be at the origin of confinement \cite{'tHooft:1977hy,Mandelstam:1974pi}. The main advantage of the Schwinger model with massless fermions (quarks) is that it is exactly soluble and can be used to investigate e.g. the role of LPM effect at strong coupling.   
Of course, for our purposes we have to generalize it to allow for the rotation of quarks in color space as they traverse the quark-gluon plasma (recently the effects of the color flow were addressed in \cite{Beraudo:2011bh}).

\begin{figure}[htbp]
\centering
\includegraphics[width=1\linewidth]{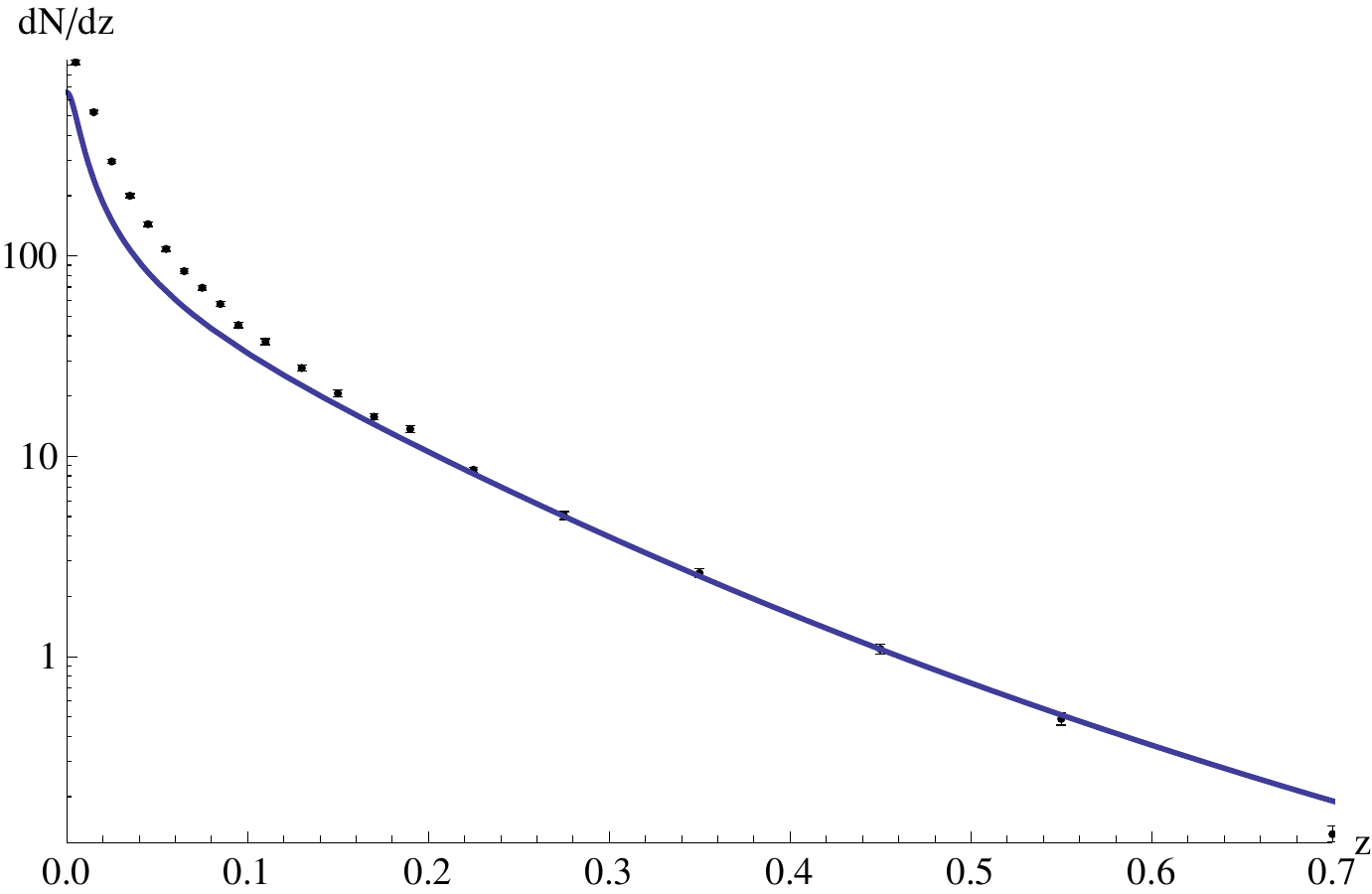} 
\caption{The spectrum of charged hadrons in $e^+e^-$ annihilation at $\sqrt{s}=201$ GeV; solid line is obtained from \eqref{eq:mdi_1}.}
\label{fig:fragmentation201GeV}
\end{figure}

The Lagrangian of QED in $1+1$ dimensions is given by 
\begin{equation}
\mathcal{L}=-\frac{1}{4}F_{\mu\nu}F^{\mu\nu}+\bar{\psi}i\gamma^\mu\partial_\mu\psi-g\bar{\psi}\gamma^\mu\psi A_\mu
\label{eq:qed}
\end{equation}
where $A_\mu$ is the $U(1)$ gauge field, $g$ is the coupling constant and $\psi$ is a Dirac spinor; $x^\mu=(t,z)$. The electromagnetic vector current is given by $
j^\mu(x)\equiv \bar{\psi}(x)\gamma^\mu\psi(x) $. 
It is well known that the theory $\eqref{eq:qed}$ admits a bosonic representation in terms of a free massive scalar field theory with a scalar field $\phi$ of mass $m$ \cite{Lowenstein:1971fc,Coleman:1975pw}:
\begin{eqnarray}
j^\mu(x)&=&\frac{1}{\sqrt{\pi}}\epsilon^{\mu\nu}\partial_\nu \phi(x) \nonumber \\
m^2&=&g^2/\pi
\label{eq:bos}
\end{eqnarray}
In \cite{:1974cks}, to describe $e^+e^-$ annihilation, an external current $j^\mu_{ext}$ of the produced quark and antiquark was added to the theory; the equation of motion becomes
\begin{equation}
(\Box+m^2)\phi(x)=-m^2\phi_{ext}(x)
\label{eq:eom}
\end{equation}
where 
\begin{equation}
j_{ext}^\mu(x)=\frac{1}{\sqrt{\pi}}\epsilon^{\mu\nu}\partial_\nu \phi_{ext}(x)
\label{eq:ext}
\end{equation}
According to  \eqref{eq:bos},  an external charge density corresponds to the scalar field 
\begin{equation}
\phi_{ext}(x)=\sqrt{\pi}\int^z{dz'j^0_{ext}(t,z')}
\label{eq:pex}
\end{equation}
If we consider as an external source the pair of quark and antiquark moving along the light cone, the charge density is given by
\begin{equation}
j_{ext}^0(x)=\delta(z-t)\theta(z)-\delta(z+t)\theta(-z)
\label{eq:cde}
\end{equation}
Given \eqref{eq:pex}, the solution to \eqref{eq:eom}, with boundary conditions $\phi(t<0,z)=0$ and $t^2>z^2$, was shown to be \cite{:1974cks}
\begin{equation}
\phi(x)=\theta(t+z)\theta(t-z)-\Delta_R(m^2,x^2)
\label{eq:sol}
\end{equation}
where $\Delta_R(m^2,x^2)$ is the retarded propagator. It is easy to check that the retarded propagator has the form 

\begin{equation}
\Delta_R(m^2,x^2)=J_0(m|x|)\theta(t+z)\theta(t-z)
\label{eq:prp}
\end{equation}
where $J_0$ is the Bessel function of the first kind and $|x|=\sqrt{t^2-z^2}$. The solution to the equation of motion now reads
\begin{equation}
\phi(x)=\theta(t+z)\theta(t-z)(1-J_0(m|x|))
\label{eq:sol_1}
\end{equation}
The (anti)kinks of the scalar field describe the production of quark-antiquark pairs in the vacuum induced by the external source. The bosonization allows us to describe this process in terms of the observed meson field $\phi$. 
Let us evaluate the momentum distribution of the produced mesons. Eq. \eqref{eq:eom} describes an interacting theory of the scalar field $\phi$ with a classical source $f(x)$.
 By using the Fourier representation of the free field and retarded propagator the momentum distribution of the created quanta is given by \begin{equation}
\frac{dN}{dp}=<0|a_p^\dagger a_p|0>=\frac{|\tilde{f}(p)|^2}{2 E_p}
\label{eq:mdi}
\end{equation}
where $\tilde{f}(p)=\int{d^2x f(x) e^{i p \cdot x}}$ and $E_p=\sqrt{p^2+m^2}$.
Let us now consider the quark and antiquark moving with a velocity $v<1$ creating the charge density
\begin{equation}
j^0(x)=\delta(z-vt)\theta(z)-\delta(z+vt)\theta(-z)
\label{eq:cde_1}
\end{equation}
Using \eqref{eq:pex} 
and \eqref{eq:mdi}, we get
\begin{equation}
\tilde{f}(p)=\sqrt{\pi}\frac{-2vm^2}{E_p^2-v^2p^2}
\label{eq:src_1}
\end{equation}
from where
\begin{equation}
\frac{dN}{dp}=2\pi\frac{v^2m^4}{E_p(E_p^2-v^2p^2)^2};
\label{eq:mdi_1}
\end{equation}
note that there is no factor of $N_c$ here since we consider hadron yield per produced jet (with a fixed color orientation).
For $v=1$, we get $dN/dp \propto 1/E_p$ familiar from the usual bremsstrahlung spectrum in $(3+1)$ dimensions. Let us define the usual fragmentation variable $z=p^h/p^{jet} \equiv p/p^{jet}$ as the fraction of jet's momentum $p^{jet}$ carried by the hadron of momentum $p^h=p$. As an application of the model, we use \eqref{eq:mdi_1} to evaluate $dN/dz$ and fit it to the data on  $e^+e^-\rightarrow {\rm hadrons}$ at $\sqrt{s}=201.7$ GeV \cite{Abbiendi:2002mj}. The result is shown in Fig. \ref{fig:fragmentation201GeV}; the fit parameters are the scalar meson mass $m$ and the matching scale $Q_0$ at which the DGLAP evolution has to be matched onto our model. From the fit we find $m=0.6$ GeV that is consistent with the PDG value for the $\sigma$ meson, and $Q_0 = 2$ GeV (the velocity $v=P_{jet}/\sqrt{P_{jet}^2 + Q_0^2}$). At small $z\leq 0.1$, our result is below the data points signaling the need for perturbative QCD evolution; however the spectrum at  $z \geq 0.15$ is reproduced reasonably well.

\begin{figure}[htbp]
\centering
\includegraphics[width=0.8\linewidth]{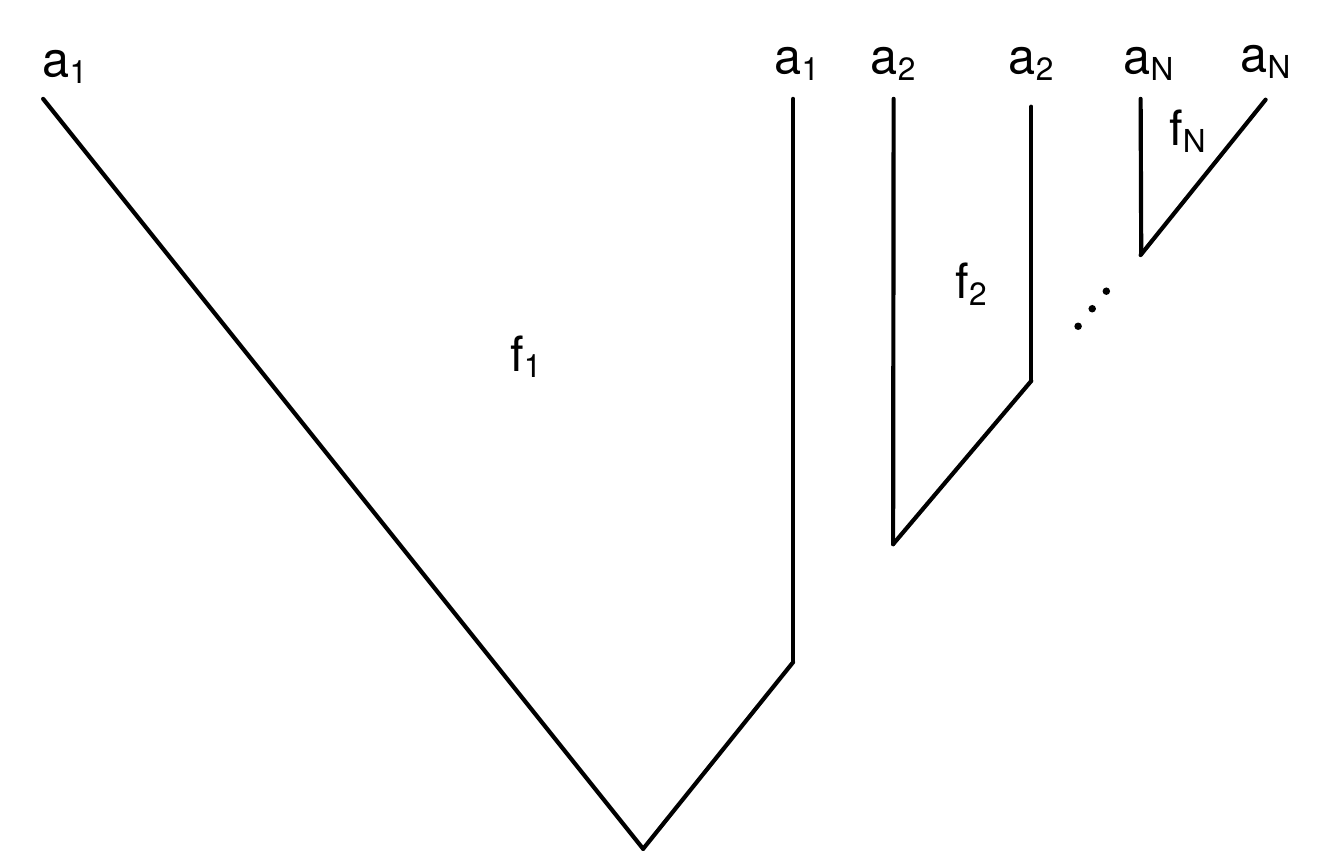} 
\caption{The color flow in the jet interactions inside the quark-gluon medium.}
\label{fig:rescattering_in_medium}
\end{figure}

Let us now extend this formalism to the case of a jet propagating through the quark-gluon matter. The quark will exchange color with the matter, rotating in color space and creating in the medium the static color sources located at coordinates $z=z_i$, see Fig. \ref{fig:rescattering_in_medium}. The different sectors in Fig. \ref{fig:rescattering_in_medium} are bounded by the sources with different orientations in color space; each of them is considered as quasi-Abelian, and at large $N_c$ the production of mesons in each sector is independent.

\begin{figure}[htbp]
\centering
\includegraphics[width=1\linewidth]{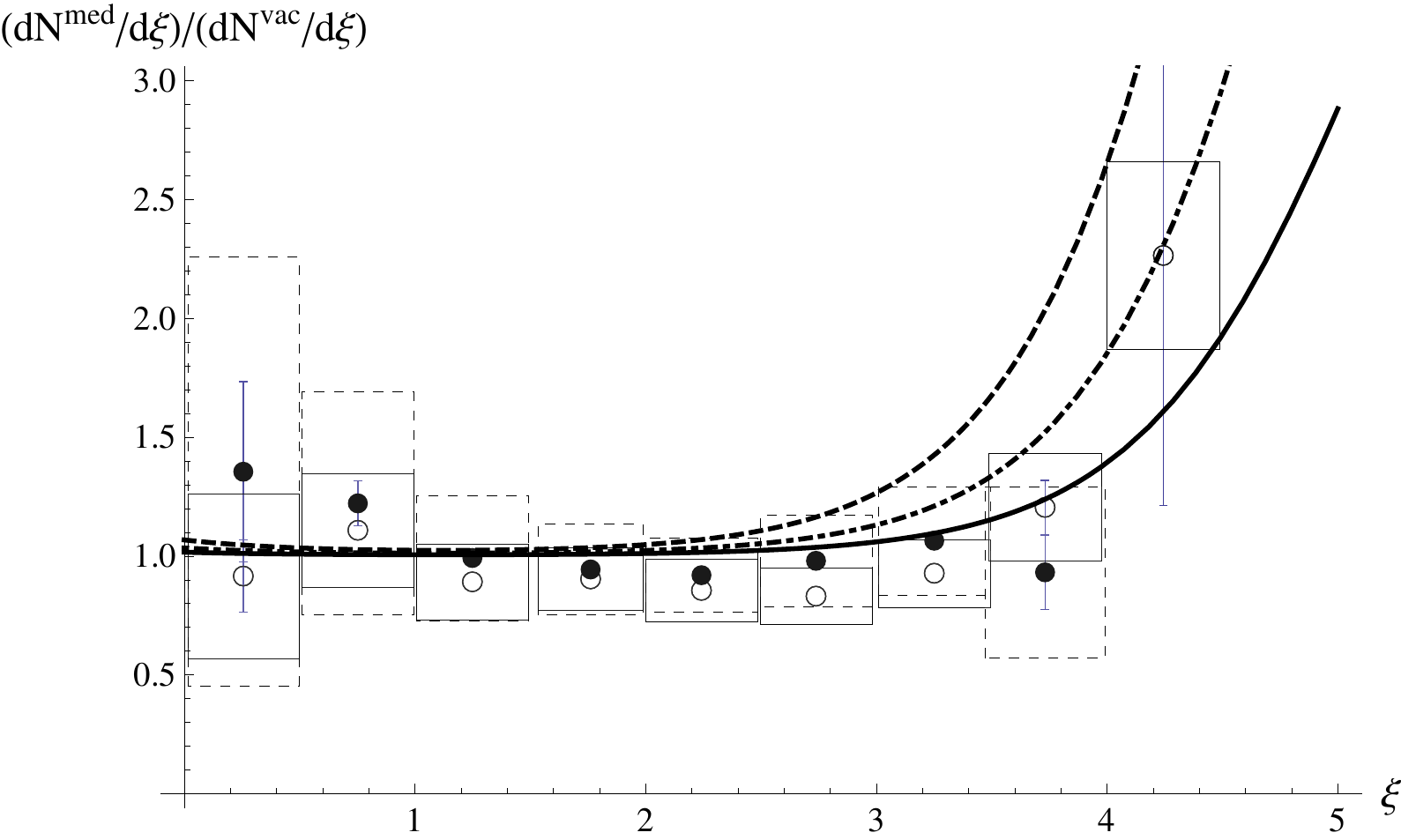} 
\caption{Ratio of in-medium to vacuum fragmentation functions. The length of the medium is fixed at $4$ fm, the jet energy is $E_{jet} = 100$ GeV. Solid line: the first scattering occurs at $t_1=1$ fm (assumed thermalization time), and subsequent scatterings occur with time spacing of $\Delta t=1/m=0.3$ fm. 
Dashed line: double scattering with $t_1=2$ fm and $t_2=4$ fm ($\Delta t = 2$ fm). Dot-dashed line:  four scatterings with $\Delta t=1$ fm, $t_1=1$ fm. Data points are from \cite{Collaboration:2011cs}. Open (filled) circles are for the leading (subleading) jet.}
\label{fig:frag_ratio120GeV}
\end{figure}

\begin{figure}[htbp]
\centering
\includegraphics[width=0.8\linewidth]{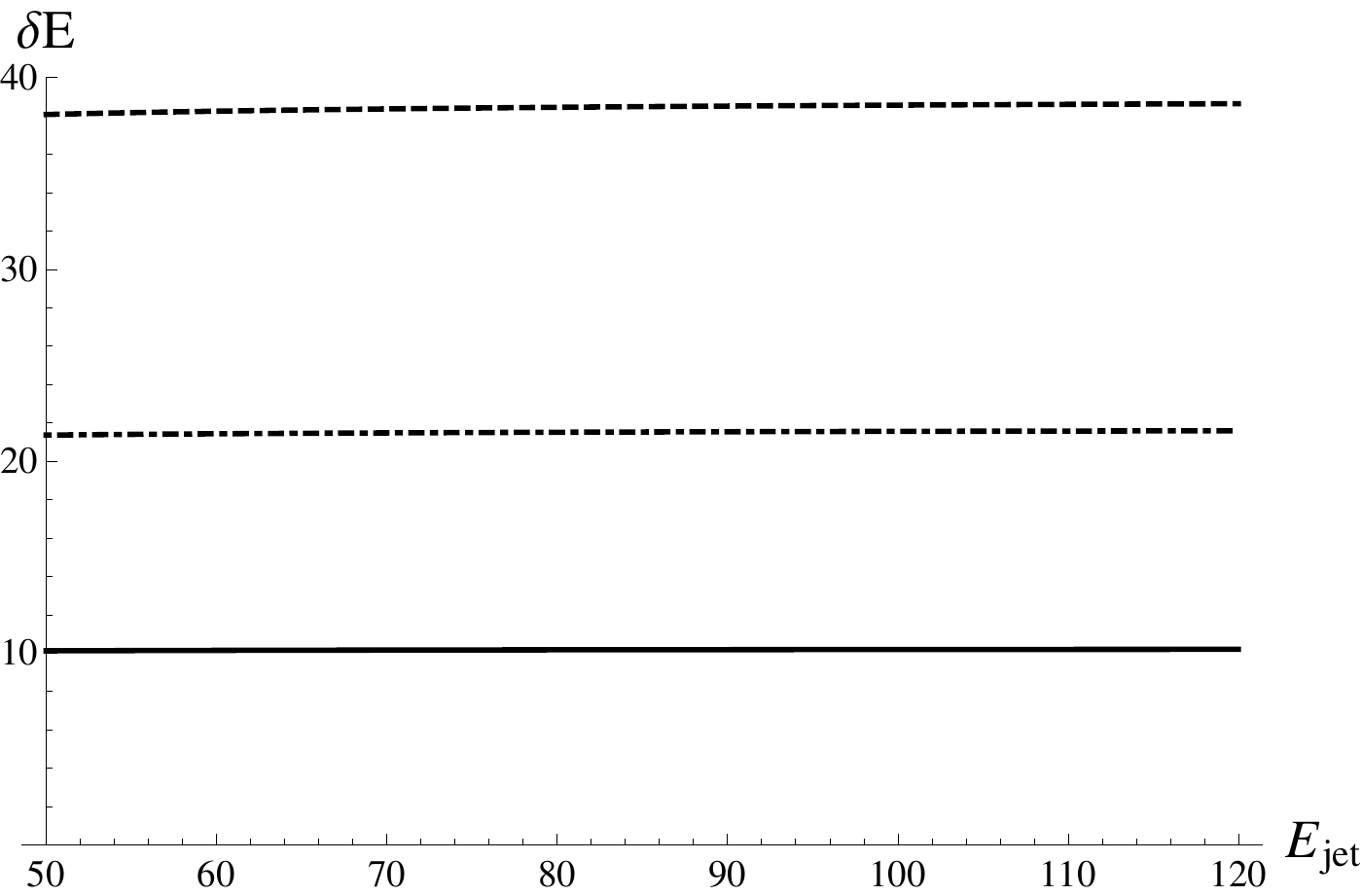} 
\caption{Energy loss as a function of jet energy. The lines correspond to the parameters in the caption of Fig. \ref{fig:frag_ratio120GeV}.}
\label{fig:EnlossvsJetEn}
\end{figure}

 We see that there are only three different types of sectors in Fig. \ref{fig:rescattering_in_medium} -- the one bounded by the quark escaping from the medium with no interactions (on the left, $f_1$), the one bounded by the quark that underwent color rotation(s) in the medium (on the right, $f_3$), and the one bounded on the sides by color static sources in the medium and a propagating quark from below (in the middle, $f_2$).  By using the methods described above, we get for the corresponding sources (we denote with $t_1$ and $t_2$ the time when the first and second scatterings in medium occur respectively)
\begin{eqnarray}
\tilde{f}_1(p)&=& \frac{-m^2v\sqrt{\pi}}{E_p-vp}\left[\frac{2}{E_p+vp}-\frac{e^{i(E_p-vp)t_1}}{Ep}\right] \nonumber \\
\tilde{f}_2(p)&=& \frac{m^2v\sqrt{\pi}}{E_p(E_p-vp)}\left[e^{i(E_p-vp)t_2}-e^{i(E_p-vp)t_1}\right] \nonumber \\ 
\tilde{f}_3(p)&=&\frac{-m^2v\sqrt{\pi}}{E_p(E_p-vp)}e^{i(E_p-vp)t_2}
\label{eq:}
\end{eqnarray}
Summing over the color orientations of different sectors (note that this does not bring in extra powers of $N_c$ in the 't Hooft limit of $N_c \to \infty$, $g^2 N_c$ fixed), we get the hadron spectrum 
\begin{equation}
\frac{dN^{med}}{dp}=\frac{1}{2E_p} |\tilde{f}(p)|^2 = \frac{1}{2E_p}(|\tilde{f}_1(p)|^2+|\tilde{f}_2(p)|^2+|\tilde{f}_3(p)|^2)
\label{eq:tcu}
\end{equation}
where we have omitted the interference between different sectors that is suppressed at large $N_c$. We can write
\begin{eqnarray} 
\frac{dN^{med}}{dp}&=&4\pi v^2\frac{m^4}{2E_p}\Bigg\{\frac{1}{(E_p^2-v^2p^2)^2}+\frac{1}{E_p^2(E_p-vp)^2} \nonumber  \\ 
	                 &-&\frac{1}{2}\frac{1}{E_p(E_p-vp)^2}\Bigg[\frac{2\cos(E_p-vp)t_1}{E_p+vp} \nonumber \\ 
	                 &+&\frac{\cos[(E_p-vp)(t_2-t_1)]}{E_p}\Bigg]  \Bigg\}
\label{eq:scattmedfin}
\end{eqnarray}
 In order to compare our result to the CMS data \cite{Collaboration:2011cs}, we use the variable $\xi=\ln\left(1/z\right)$.
Our result for the ratio of in-medium and vacuum fragmentation functions is shown in Fig.  \ref{fig:frag_ratio120GeV}. We put $N$ equally spaced scatterings between $t_1$ and $t_2$ (the distance between them corresponds to the mean free path).  One can see that the observed jet fragmentation scaling (JFS) is well reproduced if the mean free path $\lambda_{mfp}$ of the quark in the medium is short, $\lambda_{mfp} \leq m^{-1} \simeq 0.3$ fm.  
It is important to check whether the JFS in our computation stems from the absence of the energy loss -- this would contradict the experimental observations \cite{Collaboration:2011cs,Aad:2010bu,Chatrchyan:2011sx,Aamodt:2010jd}. 
The energy loss of the jet in medium is given by 
\begin{equation}
\delta E=\int_{m_h}^{E_{jet}}{dE_h E_h\left(\frac{dN^{med}}{dE_h}-\frac{dN^{vac}}{dE_h}\right)}
\label{eq:enloss}
\end{equation}
We can use \eqref{eq:enloss} to calculate the energy loss $\delta E$ as a function of jet energy $E_{jet}$. We plot this in Fig. \ref{fig:EnlossvsJetEn}; note that our treatment is valid only when $\delta E \ll E_{jet}$; for short mean free path $\lambda_{mfp} \leq 0.3$ fm this means $E_{jet} \geq 100$ GeV. The energy loss at $\lambda_{mfp} \leq 1$ fm is consistent with the values extracted from the data \cite{Collaboration:2011cs}, see \cite{CasalderreySolana:2010eh}. 
\vskip0.3cm
To summarize, we have evaluated the jet fragmentation function in QCD matter by using a non-perturbative approach based on the exactly soluble quasi-Abelian model. We find that the observed jet fragmentation scaling is reproduced for any value of the mean free path that is shorter than $0.3$ fm. Meanwhile, the induced energy loss is large and consistent with experimental observations -- this happens because the produced hadrons are abundant but possess small transverse momenta. 
Of course, our treatment has been to large extent model-dependent and can be both questioned and improved. However we 
feel that the origin of our result -- the LPM effect at strong coupling and short mean free path -- may be generic for a broad class of models.

\vskip0.1cm 

We are grateful to Wit Busza and Yuri Dokshitzer for useful discussions.
This work was supported by the U.S. Department of Energy under Contracts No.
DE-AC02-98CH10886, and DE-FG-88ER41723.

 \vfil


\end{document}